**Submicrometer tomographic resolution examined using a micro-fabricated test object**

Ryuta Mizutani [a, *], Akihisa Takeuchi [b], R. Yoshiyuki Osamura [c], Susumu Takekoshi [c], Kentaro Uesugi [b], Yoshio Suzuki [b]

[a] Department of Applied Biochemistry, School of Engineering, Tokai University, Kitakaname 1117, Hiratsuka, Kanagawa 259-1292, Japan

[b] Research and Utilization Division, JASRI/SPring-8, Kouto 1-1-1, Sayo, Hyogo 679-5198, Japan

[c] Department of Pathology, Tokai University School of Medicine, Shimokasuya 143, Isehara, Kanagawa 259-1193, Japan

*Corresponding author. Tel: +81-463-58-1211; fax: +81-463-50-2506.

E-mail address: ryuta@tokai-u.jp (R. Mizutani).

**Abstract**

To estimate the spatial resolution of microtomographs, a test object on the submicrometer scale was prepared by focused ion beam milling and subjected to microtomographic analysis. Since human tissues are composed of cells and extracellular matrices with micrometer and submicrometer structures, it is important to investigate the three-dimensional spatial resolution of microtomographs used to visualize microstructures of human tissues. The resolutions along the direction within the tomographic slice plane (in-plane resolution) and perpendicular to it (through-plane resolution) were determined from the modulation transfer function of square-wave patterns. The in-plane resolution was estimated to be 1.2 um from the modulation transfer function of the non-zoom image. In contrast, the zoom image gave the in-plane resolution of 0.8 um. This in-plane resolution is comparable to the through-plane resolution, which was estimated to be 0.8 um. Although the two-dimensional radiographs were taken with the pixel width of half the x-ray optics resolution, these three-dimensional resolution analyses indicated that the zoom reconstruction should be performed to achieve the in-plane resolution comparable to the x-ray optics resolution. The submicrometer three-dimensional analysis was applied in the structural study of human cerebral tissue stained with high-Z elements and the obtained tomograms revealed that the microtomographic analysis allows visualization of the subcellular structures of the cerebral tissue.

# 1. Introduction

The application of synchrotron radiation to microtomographic analysis has allowed high-resolution three-dimensional analysis (Bonse and Busch, 1996; Salome et al., 1999; Uesugi et al., 2001; Takeuchi et al., 2002). Microtomographic analyses have revealed the three-dimensional structures of human tissues including bones (Bonse et al., 1994; Peyrin et al., 2001) and cerebral cortex (Mizutani et al., 2008a). Biological tissues are composed of cells and extracellular matrices with typical dimensions on the micrometer to submicrometer scales. Subcellular structures such as the nucleus are essential for biological functions. Therefore, the three-dimensional spatial resolution of computed microtomographs (micro-CTs) should be examined to visualize the three-dimensional microstructure of human tissues.

The modulation transfer function (MTF) of clinical CTs has been measured using a flat surface slanted with respect to the reconstruction matrix (Judy, 1976) or a slit phantom consisting of metal foil and tissue equivalent slabs (Boone, 2001). The MTF can also be calculated from the variance and amplitude of the square-wave pattern (Droege and Rzeszotarski, 1985). Although the MTFs of clinical CTs can be estimated from these test objects on the millimeter scale, MTF estimation of microtomographs should be performed by using test objects with micrometer precision. It has been reported that an optical quality ball can be used as a test object to measure the MTF of a microtomograph (Seifert and Flynn, 2002). We have recently reported the fabrication of a three-dimensional test pattern on the micrometer scale by focused ion beam (FIB) milling (Mizutani et al., 2008c). Those test objects revealed micrometer resolutions of laboratory and synchrotron radiation microtomographs. However, the three-dimensional submicrometer resolution achieved by recent microtomographs has not been examined because a three-dimensional test object appropriate for estimating the submicrometer resolution has not been prepared.

In the study reported here, a three-dimensional test object on the submicrometer scale was fabricated by FIB milling. The resolutions along the directions within the tomographic slice plane (hereafter called in-plane resolution) and perpendicular to it (through-plane resolution) were estimated by using the test object. Although the two-dimensional radiographs were taken with the pixel width of half the x-ray optics resolution, the three-dimensional resolution analyses revealed that the zoom reconstruction should be performed to achieve submicrometer resolution. The microtomographic analysis was then applied to the structural analysis of human cerebral tissue to visualize the cellular and subcellular structures of neurons.

# 2. Materials and Methods

## 2.1. Submicrometer test object

A three-dimensional test object for the submicrometer resolution estimation was fabricated by using an FIB apparatus (FB-2000, Hitachi High-Technologies, Japan) operated at 30 kV. An aluminum wire with a diameter of 250 μm was attached to the sample holder and subjected to gallium ion beam milling. In the previous report, square-wave patterns were prepared directly on the wire surface (Mizutani et al., 2008c). However, the curvature of the wire surface cannot be ignored in the fabrication of a submicrometer test object. Therefore, a flat plane with roughness less than 50 nm was prepared before carving the square-wave pattern. A gallium beam of 4.5 nA was used for the rough abrasion of a 50-μm wide surface of the wire and a 1.7 nA beam for finishing the flat surface. A series of square wells was carved on this flat surface. The pitches of the resultant square-wave patterns were 2.0, 1.6, 1.2, 1.0, 0.8, 0.6, 0.5, and 0.4 μm. Each pattern had a 50% duty cycle, *i.e.*, 0.5-μm well and 0.5-μm interval for a 1.0-μm pitch. A gallium beam of 0.33 nA was used for carving the 2.0, 1.6, and 1.2 μm patterns and a 0.08 nA beam for the 1.0, 0.8, 0.6, 0.5, and 0.4 μm patterns, giving rectangular wells with a typical depth of 4 μm. A secondary electron image of the obtained test object is shown in Fig. 1. The test object was then recovered and mounted on a stainless steel pin by using epoxy glue. Air exposure had no effect on the pattern structure.

*2.2. Human cerebral tissue*

Cerebral tissues with dimensions of about 0.3 mm × 0.3 mm × 3 mm were prepared as described previously (Mizutani et al., 2008a) and subjected to Bodian impregnation by using silver nitrate and hydrogen tetrachloroaurate as stain dyes (Mizutani et al., 2007). Anatomical analysis found no abnormality in the brain tissue. Human samples were obtained with consent using protocols approved by the Clinical Research Review Board of Tokai University Hospital.

The stained tissues were sequentially immersed in ethanol, *n*-butylglycidyl ether, and Petropoxy 154 epoxy resin (Burnham Petrographics, USA). The samples were transferred into a borosilicate glass capillary (W. Müller Glas, Germany) filled with epoxy resin. The outer diameter of the capillary was 0.7 mm. The capillaries were then incubated at 90ºC for 16 hr to cure the epoxy resin.

*2.3. Microtomographic analysis*

The microtomographic analysis was performed at the BL20XU beamline (Suzuki et al., 2004) of SPring-8. The steel pin with the test object or the glass capillary containing the cerebral tissue was mounted on the goniometer head of the microtomograph using a brass fitting designed for the pin-hold sample. Transmission radiographs were recorded using a CCD-based x-ray imaging detector (AA50 and C4880-41S, Hamamatsu Photonics, Japan) and 12.000-keV x-rays. Each radiograph was taken by averaging pixels in 2 × 2 bins. The number of detector pixels after binning was 2000 in the horizontal direction perpendicular to the sample rotation axis and 1312 along the vertical axis. These radiographs with the effective pixel size of 0.50 μm × 0.50 μm were acquired in a parallel-beam geometry using the synchrotron radiation, giving the same pixel size at the sample position. Hence, the field of view was 1000 μm × 656 μm. The images of this area were acquired with a rotation step of 0.10º and exposure time of 300 ms per image. The data acquisition conditions are summarized in table 1.

*2.4. Microtomographic reconstruction*

The microtomographic reconstruction was performed by the Donner algorithm (Huesman et al., 1977). The convolution back-projection method using a Hann-window filter (Chesler and Riederer, 1975; Huesman et al., 1977) was used for the reconstruction calculation. The non-zoom reconstruction was performed with a frequency cutoff at half the total bandwidth. For the zoom reconstruction, the horizontal pixel strip extracted from the two-dimensional radiograph was subjected to double or quadruple linear interpolation, as described previously (Mizutani et al., 2008c). The double interpolation was performed by placing one additional pixel between observed pixels, and the quadruple interpolation by placing three pixels. Each interpolated strip was subjected to a Fourier transformation. The filter functions with a frequency cutoff at half the total bandwidth were applied to the Fourier transform of the interpolated strips. Then the inverse Fourier transformation and back-projection calculations were performed to obtain the microtomogram. These reconstruction calculations were performed by using the program RecView (available from http://www.el.u-tokai.ac.jp/ryuta/) accelerated with CUDA parallel-computing processors.

These tomographic reconstructions were performed taking into account the monochromatic x-ray transmittance at each pixel in the two-dimensional image and the voxel size of the resultant three-dimensional image. The voxel values of the obtained tomograms are calculated as a linear absorption coefficient (LAC) at the observed x-ray energy of 12 keV. These observed absorption coefficients have been calibrated with standard materials (Tsuchiyama et al., 2005). The LACs used in this study were calculated in accordance with this calibration study.

The in-plane resolution within the tomographic slice should be affected by the reconstruction calculation, while the through-plane resolution along the rotation axis is largely independent of the tomographic reconstruction. Therefore, the in-plane and through-plane resolutions were both estimated. The variance and amplitude of the square-wave patterns in the in-plane and through-plane directions were analyzed to obtain the MTFs (Droege and Rzeszotarski, 1984, 1985). The in-plane resolution was examined from the non-, double- and quadruple-zoom images of the pattern perpendicular to the sample rotation axis. The through-plane resolution was examined from the non-zoom images of the pattern along the axis.

## 3. Results and Discussion
### 3.1. Resolution estimation

Microtomograms of in-plane patterns obtained by the non-zoom and quadruple-zoom reconstruction are shown in Fig. 2. The zoom reconstruction resolved the in-plane pattern up to a pitch of 1.0 μm, whereas the non-zoom reconstruction resolved only 2.0 and 1.6 μm patterns. The x-ray optics resolution has been estimated to be 1 μm (Uesugi et al., 2001). Although the two-dimensional radiographs were taken with the pixel width of 0.5 μm corresponding to half the x-ray optics resolution, the zoom reconstruction should be performed to resolve the 1.0 μm pattern.

The spatial resolution was estimated from the microtomographic image of the square-wave pattern. The variance and amplitude of the pattern were used for calculating the MTF values (Droege and Rzeszotarski, 1984, 1985). The obtained MTF in each direction is plotted in Fig. 3. The non-zoom in-plane resolution at 5% MTF was estimated to be approximately 1.2 μm, which is consistent with the value estimated from microtomograms of amorphous silica grains (Uesugi et al., 2001). In contrast, the double-zoom MTFs gave the in-plane resolution of 0.8 μm. This in-plane resolution is comparable to the through-plane resolution, which was estimated to be 0.8 μm at 5% MTF.

In many microtomographic analysis cases, the pixel width of the two-dimensional image is set to around half the x-ray optics resolution in order to minimize the amount of raw data, which typically reaches over $10^{10}$ bytes. This Nyquist criterion has also been applied to the image sampling in clinical CTs (Lin et al., 1993; Goldman, 2007). However, the spatial resolution is the most important concern in microtomographic analysis. From the sampling theorem, $N$-dimensional images should be taken with a pixel width of less than $1/2\sqrt{N}$ times the spatial resolution (Pawley, 2006). Therefore, sampling at half the spatial resolution is insufficient for microtomographic analysis of three-dimensional objects. The three-dimensional image should be digitized with a pixel width of $1/2\sqrt{3}$ times the spatial resolution. This pixel width can be achieved by the double-zoom reconstruction, which gives a pixel width of 1/4 of the spatial resolution. Therefore, zoom reconstruction can retrieve the high-resolution information resolved by the x-ray optics.

When the sampling theorem in $N$-dimensional space is taken into account, the pixel width should be 0.23 μm in order to resolve a 0.8-μm structure. In this study, the two-dimensional images with the pixel width of 0.5 μm were taken by averaging 0.25-μm pixels in 2 × 2 bins. The spatial resolution can be improved by taking images with a pixel width of 0.25 μm by 1 × 1 binning. Finer resolution can also be achieved by shifting the detector by 0.25 μm while the images are taken with the 2 × 2 averaging. If data acquisition by the 1 × 1 binning or detector-shift method is achieved, submicrometer structures should be resolvable without artificial zoom reconstruction.

It has been reported that the MTF of a microtomograph can be estimated by imaging a ruby sphere with sphericity of 0.64 μm (Seifert and Flynn, 2002). However, this method gives an overall average of spatial resolutions along all directions because radial profiles of the sphere surface are projected into a one-dimensional space. The present results indicate that the MTFs in the in-plane and through-plane directions were different and should be estimated separately. This anisotropy of the spatial resolution can be estimated by using the square-wave pattern. Nanometer precision of the FIB fabrication is also essential for estimating the submicrometer resolution.

*3.2. Human cerebral tissue*

The recent application of microtomographic analysis to biological objects has resolved three-dimensional structures at the micrometer to submicrometer scales (Bonse et al., 1994; Peyrin et al., 2001; Mizutani et al., 2007). In conjunction with metal microcontrasting, microtomographic analysis has revealed cellular and subcellular structures of soft tissues including the human cerebral cortex (Mizutani et al., 2008a) and capillary vessels (Mizutani et al., 2008b).

Structures of human cerebral cortex stained with Bodian impregnation are shown in Fig. 4. Reduced-silver Bodian impregnation is a conventional method used for the light microscopic observation of neural tissues (e.g. Tyrer et al., 2000). This method stains neuropils with gold particle (Mizutani et al., 2007). A histological image of the cerebral tissue adjacent to the sample for the microtomographic analysis (Fig. 4a) indicated that the tissue was intensely stained near the tissue surface but weakly in the interior part of the tissue block. The Bodian impregnation has been developed to stain thin slices rather than block tissues. Therefore, the staining was not suitable to visualize the entire tissue blocks by light microscopy.

However, the intensely stained layer visualized by light microscopy was not observed in the microtomographic image. Microtomographic slices of the three-dimensional structure of the cerebral tissue are shown in Fig. 4b and 4c. Although the stained contrast was insufficient for visualizing the entire network, spherical structures along with some neuropils can be seen in the microtomograms. These spherical structures have diameters of approximately 5 μm and are compatible with cellular nuclei of smaller neural cells. Within these spherical structures, even denser granules corresponding to nucleoli were observed. These fine structures were better resolved in the zoom reconstruction rather than in the non-zoom reconstruction (Fig. 4b and 4c).

## 4. Conclusions

The spatial resolution is the fundamental parameter in structural analysis. The three-dimensional test pattern on the submicrometer scale allowed the quantitative evaluation of the spatial resolution of the high-resolution microtomographs. Although the two-dimensional radiographs were taken with the pixel width of half the x-ray optics resolution, the three-dimensional resolution analyses revealed that the zoom reconstruction should be performed to achieve the in-plane resolution comparable to the x-ray optics resolution.

The submicrometer microtomographic analysis was applied in the structural study of human cerebral tissue stained with high-Z elements. The obtained tomograms indicated that the microtomographic analysis allows visualization of the subcellular structures of the cerebral tissue. Although the staining procedure should be improved to visualize the entire network, the neuronal structure that can be visualized by the three-dimensional analysis is essential for cerebral functions.

**Acknowledgements**

We thank Yasuo Miyamoto, Technical Service Coordination Office, Tokai University, for helpful assistance with FIB milling. We also thank Noboru Kawabe, Teaching and Research Support Center, Tokai University School of Medicine, for histological analysis. The synchrotron radiation experiments were performed at SPring-8 with the approval of the Japan Synchrotron Radiation Research Institute (JASRI) (proposal nos. 2006B1716, 2007A1844, 2007B1102, and 2008B1261).

**Figures**

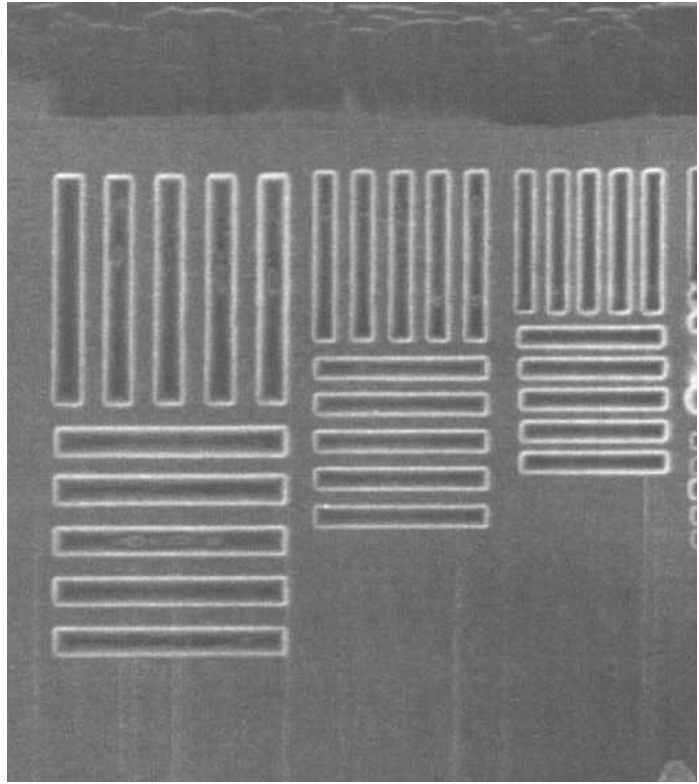

**Fig. 1.** FIB secondary electron image of the three-dimensional test object. The wire axis is horizontal. In-plane and through-plane patterns with pitches of 1.6, 1.2, and 1.0 μm are presented.

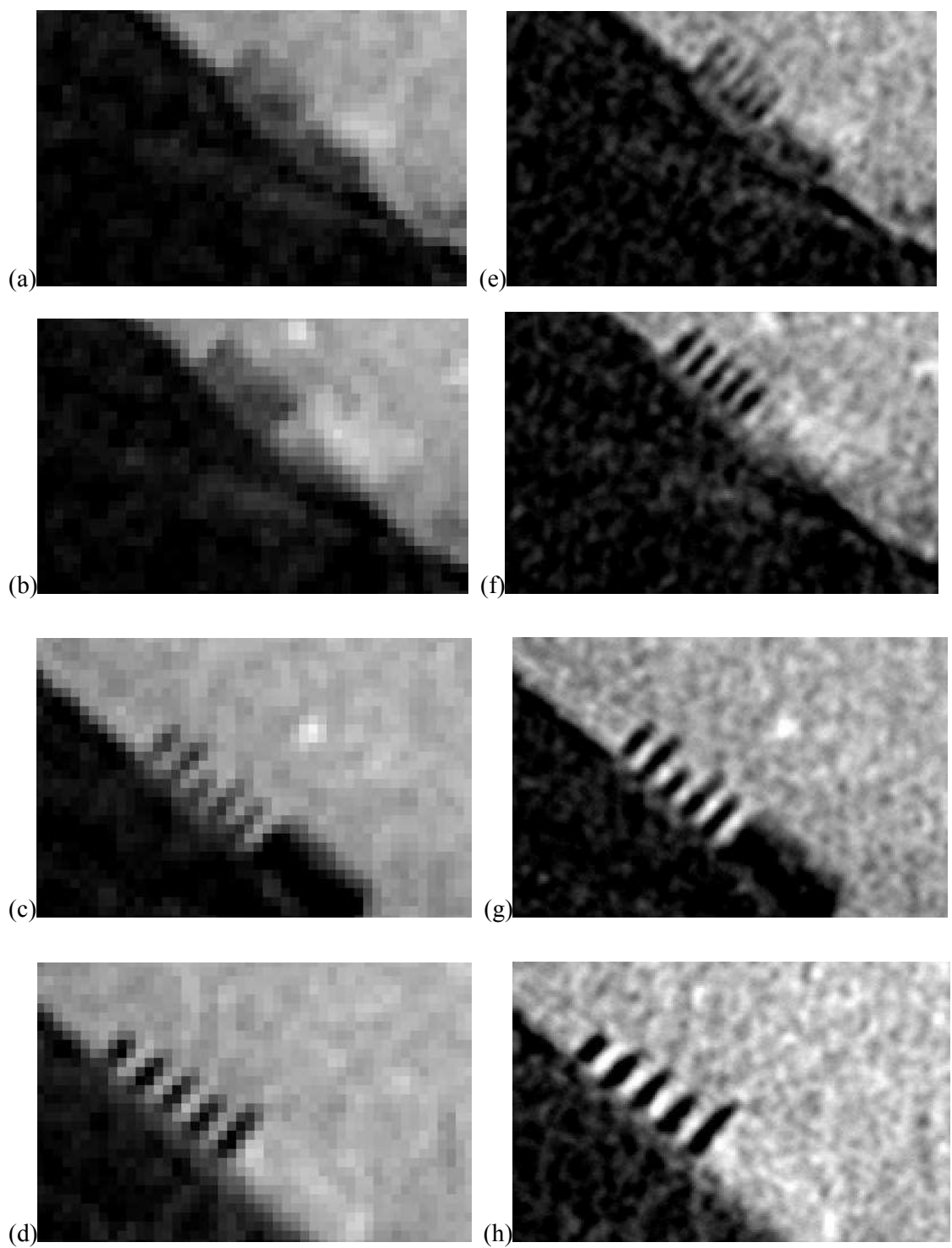

**Fig. 2.** Microtomograms of in-plane patterns obtained by non-zoom (a–d) and quadruple-zoom (e–h) reconstruction. Pattern with pitches of 1.0 µm (a, e), 1.2 µm (b, f), 1.6 µm (c, g), and 2.0 µm (d, h) are shown. LACs are shown in gray scale from 0 cm$^{-1}$ (black) to 50 cm$^{-1}$ (white).

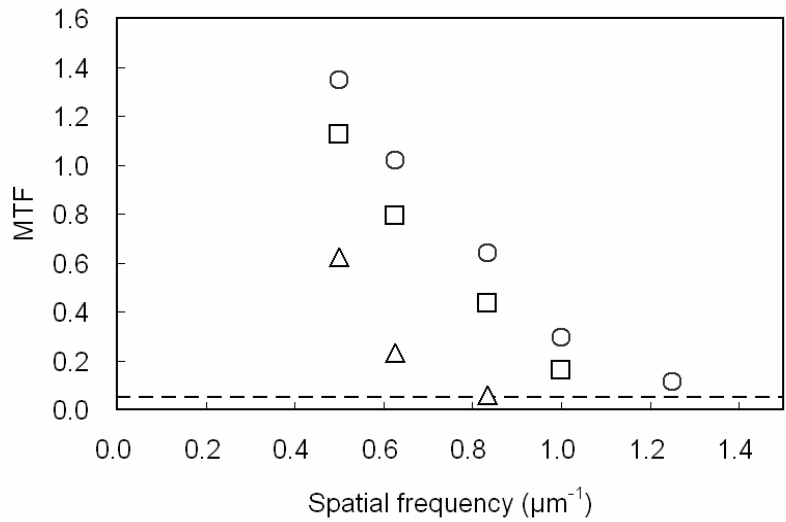

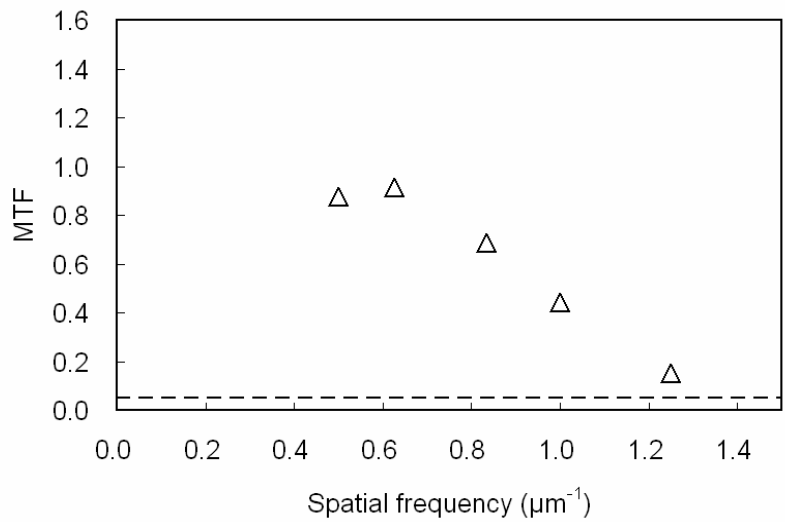

**Fig. 3.** MTFs calculated from square-wave patterns along the in-plane (a) and through-plane (b) directions. MTFs of non-, double-, and quadruple-zoom reconstructions are plotted with triangles, squares, and circles, respectively. Dashed lines represent the 5% MTF level.

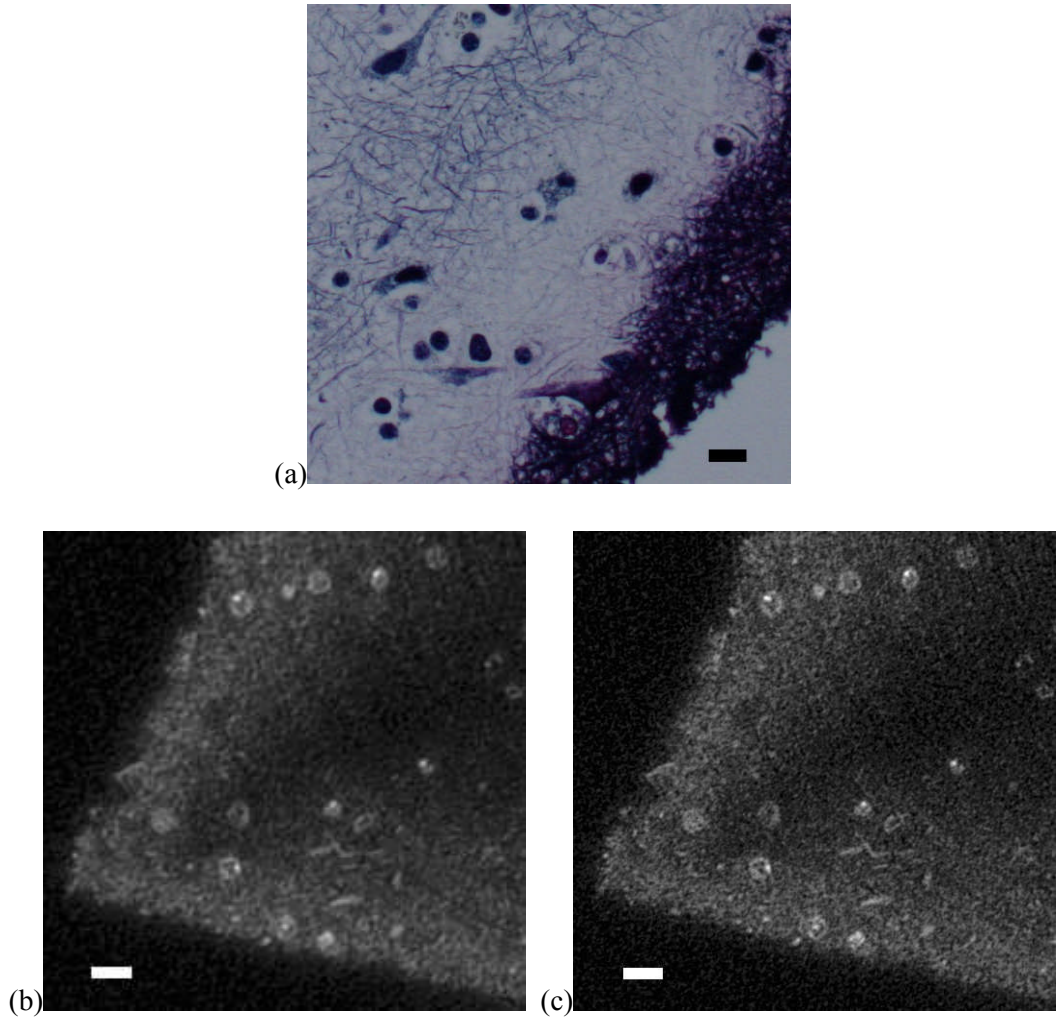

**Fig. 4.** Microstructure of human cerebral cortex. Histological image of a 3-μm section adjacent to the sample for the microtomographic analysis is shown in panel (a). Microtomograms were obtained by non-zoom (b) and double-zoom (c) reconstruction. LACs are shown in gray scale from 0 cm$^{-1}$ (black) to 50 cm$^{-1}$ (white). Scale bars: 10 μm.